\documentclass{tMPH2e}

\usepackage{color}

\newcommand{\dd}{{\rm{d}}}
\newcommand{\id}{{\rm{id}}}
\newcommand{\ee}{{\rm{e}}}
\newcommand{\ii}{{\rm{i}}}
\newcommand{\rr}{\bm{r}}

\newcommand{\nut}{z}

\newcommand{\beq}{\begin{equation}}
\newcommand{\eeq}{\end{equation}}

\newcommand\beqa{\begin{eqnarray}}
\newcommand\eeqa{\end{eqnarray}}
\newcommand{\nn}{\nonumber\\}

\def\bal#1\eal{\begin{align}#1\end{align}}

\begin{document}

\doi{10.1080/0026897YYxxxxxxxx}
 \issn{1362�3028}
\issnp{0026�8976}
\jvol{00}
\jnum{00} \jyear{2015} 

\title{{\itshape Multi-particle critical correlations}}

\author{Andr\'es Santos$^{a}$$^{\ast}$\thanks{$^\ast$Corresponding author. Email: andres@unex.es} and Jaros{\l}aw Piasecki$^{b}$
\\\vspace{6pt}  $^{a}${\em{Departamento de F\'{\i}sica and Instituto de Computaci\'on Cient\'ifica Avanzada (ICCAEx), Universidad de
Extremadura,  E-06071 Badajoz, Spain}};\\
$^{b}${\em{Institute of Theoretical Physics, Faculty of Physics, University of Warsaw, Pasteura 5, 02-093 Warsaw, Poland}}}

\maketitle

\begin{abstract}
We study the role of multi-particle spatial correlations in the appearance of a liquid--vapour critical point. Our analysis is based on the exact infinite hierarchy of equations relating spatial integrals of $(k+1)$-particle correlations to the $k$-particle ones and their derivatives with respect to the density. Critical exponents corresponding to generalised compressibility equations resulting from the hierarchy are shown to grow linearly with the order of correlations.
We prove that the critical behaviour requires taking into account correlation functions of arbitrary order. It is however only a necessary condition. Indeed, approximate closures of the hierarchy obtained by expressing higher order correlations in terms of lower order ones (Kirkwood's superposition approximation and its generalisations) turn out to be inconsistent with the critical behaviour.

\bigskip

\begin{keywords} spatial correlations; critical point; generalised compressibility equations
\end{keywords}\bigskip

\end{abstract}

\section{Introduction}
\label{sec1}

Phase transitions are accompanied by qualitative changes in the behaviour of correlation functions. In the case of an approach toward the liquid--vapour critical point along the critical isotherm the monotonic exponentially damped decay of pair correlations between the density fluctuations turns into a power-like non-integrable decay leading to a divergence of the isothermal compressibility. This important relation between the behaviour of correlations and the appearance of singularities in the thermodynamic functions suggests that also higher order correlations carry a similar information. It is the aim of this study to elucidate the fundamental role of multi-particle correlations for the appearance of a critical point and the related singularities.

In order to study this question we use the infinite hierarchy of equations derived by Baxter \cite{B64}. The hierarchy relates the spatial integrals of $(k+1)$-particle correlation functions to $k$-particle correlations and their derivatives with respect to the density, for $k=2,3, \ldots$ (Section \ref{sec2}). Our choice is dictated by the fact that the hierarchy is valid in any dimension and for any potential of interaction. It reflects thus the essential structure of the equilibrium Gibbs ensemble.

There are two kinds of questions which we study and answer in the present paper. The first concerns the evaluation of critical exponents characterising the divergence of spatial integrals of multi-particle correlation functions. The discussion of this topic is presented in Section \ref{sec4}.

The second type of problems focuses on the criteria for the occurrence of a critical point when an approximate closure of the hierarchy is adopted. For instance, we show that it is necessary to take into account correlations of arbitrary order (Section \ref{sec3}). All multi-particle correlations play thus a crucial role in the appearance of a critical point.

A related question concerns the possibility of predicting a critical point within an approximation which expresses higher order correlation functions in terms of lower order ones. A well-known example is the Kirkwood superposition (KSA) approximation discussed in Section \ref{sec5}. It turns out that this approximation applied to the exact hierarchy yields a closed equation for the pair correlation function inconsistent with a critical behaviour \cite{PSK13}. In Section \ref{sec6} we present arguments suggesting the impossibility of describing criticality even within generalised superposition approximations (GSA) where higher order correlation functions are approximated in terms of the lower order ones. The paper ends with concluding comments in Section \ref{sec7}.

\section{Multi-particle correlation functions}
\label{sec2}
Let us consider a fluid in equilibrium made of $N$ particles in a volume $V$. The (reduced) $k$-particle density $n_k(\rr_1,\rr_2,\ldots,\rr_k)$ is defined (in the canonical ensemble) as \cite{HM06}
\beq
n_k(\rr_1,\rr_2,\ldots,\rr_k)=\lim_\infty\frac{N!}{Q_N (N-k)!}\int_V\dd \rr_{k+1}\cdots \int_V\dd \rr_{N}\,\ee^{-\beta\Phi_N(\rr^N)},
\label{nk}
\eeq
where $\beta=1/k_BT$ is the inverse temperature parameter, $\Phi_N(\rr^N)$ is the potential energy (with the shorthand notation $\rr^N\equiv \{\rr_1,\rr_2,\ldots,\rr_N\}$) and
\beq
Q_N=\lim_\infty\int_V\dd \rr_{1}\cdots \int_V\dd \rr_{N}\,\ee^{-\beta\Phi_N(\rr^N)}
\label{QN}
\eeq
is the configuration integral. In Equations \eqref{nk} and \eqref{QN},
\beq
\lim_\infty\equiv \lim_{V\to\infty,N/V=n=\text{const}}
\eeq
denotes the thermodynamic limit.
Invariance of the potential energy $\Phi_N(\rr^N)$ under spatial translation implies that the same property holds for the particle densities, i.e., $n_k(\rr_1,\rr_2,\ldots,\rr_k)=n_k(\rr_1+{\bm{a}},\rr_2+{\bm{a}},\ldots,\rr_k+{\bm{a}})$ with an arbitrary vector ${\bm{a}}$.

The particle densities $n_k$ depend on the temperature $T$ and on the number density $n$ of the system. In 1964, Baxter \cite{B64} proved an elegant hierarchy relating the derivative of $n_k$ with respect to density with an integral of $n_{k+1}$, namely
\beq
\left(\chi_T n\frac{\partial}{\partial n}-k\right)n_k(1,\ldots,k)=\int\dd (k+1)\,\left[n_{k+1}(1,\ldots, k, k+1)-nn_{k}(1,\ldots, k)\right],
\label{0}
\eeq
where
\beq
\chi_T(n,T)=k_BT \left(\frac{\partial n}{\partial p}\right)_T
\label{chi}
\eeq
is the isothermal compressibility multiplied by $n k_BT$, $p$ being the pressure. In Eq.\ \eqref{0} the shorthand notations $n_k(1,\ldots,k)\equiv n_k(\rr_1,\ldots,\rr_k)$ and $\dd (k+1)\equiv \dd \rr_{k+1}$ have been introduced.

In an ideal gas, $\Phi_N(\rr^N)=0$ and thus $n_k^\id= n^k$. In that trivial case, Eq.\ \eqref{0} yields $\chi_T^\id=1$, as expected. In general, however, the interactions among particles induce spatial correlations and therefore $n_k\neq  n^k$. Consequently, one can introduce the $k$-particle distribution function $g_k(1,2,\ldots,k)$ as \cite{HM06}
\beq
n_k(1,2,\ldots,k)=n^kg_k(1,2,\ldots,k).
\eeq
It is also convenient to define the correlation functions $h_k$ by the cluster expansion \cite{UHK63}
\begin{subequations}
\bal
g_1(1)=&h_1(1)=1,
\\
g_2(1,2)=&h_1(1)h_1(2)+h_2(1,2),
\label{29}
\\
g_3(1,2,3)=&h_1(1)h_1(2)h_1(3)+\mathbf{3} h_1(1)h_2(2,3)+h_3(1,2,3),
\label{30}
\\
g_4(1,2,3,4)=&h_1(1)h_1(2)h_1(3)h_1(4)+\mathbf{6}h_1(1)h_1(2)h_2(3,4)+\mathbf{3}h_2(1,2)h_2(3,4)\nn
&+\mathbf{4}h_1(1)h_3(2,3,4)+h_4(1,2,3,4),
\label{31}
\eal
\end{subequations}
and so on. Here, a numerical coefficient in boldface means that there exist a number of analogous terms (except for
particle labelling) to the indicated canonical term. Note that, even though  the one-particle correlation function $h_1$ is equal to unity, it is included in Equations \eqref{29}--\eqref{31} to make the general cluster structure clear.
Expressed in terms of the correlation functions, the hierarchy \eqref{0} becomes \cite{B64}
\beq
\left(\chi_T n\frac{\partial}{\partial n}-k\right)n^kh_k(1,\ldots,k)=n^{k+1}\int\dd (k+1)\,h_{k+1}(1,\ldots,k, k+1).
\label{0b}
\eeq
In particular, setting $k=1$ in Eq.\ \eqref{0b}, we obtain
\beq
\chi_T-1=n\int\dd 2\,h_2(1,2).
\label{compress_eqn}
\eeq
This is nothing else but the well-known compressibility equation, conventionally derived within the grand canonical ensemble \cite{HM06}.

Our main goal is to exploit Eq.\ \eqref{0b} for $k\geq 2$. To that end, let us define the correlation integrals
\beq
I_k(n,T)\equiv n^k\int \dd2\cdots\int \dd k\, h_k(1,2,\ldots,k).
\label{1}
\eeq
Note that, because of translational invariance, $I_k$ is independent  of $\rr_1$. Integrating over the coordinates of particles $2,\ldots,k$ in both sides of Eq.\ \eqref{0b}, one has
\beq
I_{k+1}=\mathcal{L}_kI_k,
\label{2}
\eeq
where the operator $\mathcal{L}_k$ is
\beq
\mathcal{L}_k\equiv \chi_T n\partial_n -k.
\eeq
By iteration,
\beq
I_{k+1}=\mathcal{L}_k\mathcal{L}_{k-1}\cdots \mathcal{L}_2n(\chi_T-1),\quad (k\geq 2).
\label{16}
\eeq
Equations \eqref{2} and, equivalently, \eqref{16} define \emph{generalised} compressibility equations.

Equation \eqref{16} can be written in an alternative equivalent form. This follows immediately from the identity
\bal
\chi_T n\frac{\partial }{\partial n } = &k_B T n \frac{\partial n}{\partial p }\frac{\partial }{\partial n }
\nn
 =& k_B T \frac{\partial }{\partial \mu  }=z \frac{\partial }{\partial z  },
\eal
where all partial derivatives are calculated at constant temperature, $\mu $ denotes the chemical potential, $z=\ee^{\beta\mu}$ is the fugacity and the thermodynamic relation $n={\partial p}/{\partial \mu  }$ has been used.
Thus, Eq.\ \eqref{16} becomes
\beq
I_{k+1}=\left(z\partial_z-k\right)\cdots\left(z\partial_z-2\right)\left(z\partial_z-1\right)n.
\label{16b}
\eeq
The formal expansion of density as a series in powers of fugacity is \cite{H56}
\beq
n(T,z)=\sum_{\ell=1}^\infty {c_\ell}(T)\nut^\ell,
\label{n(z)}
\eeq
where each coefficient $c_\ell$ is proportional to the sum of all the cluster diagrams (reducible and irreducible) with $\ell$ points. Inserting Eq.\ \eqref{n(z)} into Eq.\ \eqref{16b} one can easily obtain
\beq
I_{k+1}(T,z)=\sum_{\ell=k+1}^\infty \frac{(\ell-1)!}{(\ell-k-1)!}{c_\ell}(T)\nut^\ell.
\eeq
Therefore, seen as a function of temperature and fugacity (grand canonical ensemble), the correlation integral $I_{k+1}$ is of order $z^{k+1}$.

To illustrate the importance of the existence of multi-particle correlations as measured by the functions $h_k$, in the next section we analyse the solution of the generalised compressibility Equation \eqref{16b} resulting from the approximate closure $I_{k+1}=0$.

\section{Solution of the generalised compressibility equation by neglecting correlations}
\label{sec3}

As said above, if one neglects pair correlations ($h_2=0$), Eq.\ \eqref{compress_eqn} yields the ideal-gas result $\chi_T^\id=1$. An interesting problem is to find out what happens if $h_2,\ldots,h_{k_0}$ are retained but $({k_0}+1)$-particle correlations are neglected, i.e., $h_{{k_0}+1}=0$. In that case,  Eq.\ \eqref{16b} with $I_{{k_0}+1}=0$ becomes a linear equation whose general solution is  \emph{polynomial},
  \beq
 n(T,z)=\sum_{\ell=1}^{k_0} {c_\ell}(T)\nut^\ell.
\label{A6a}
\eeq
Thus, the exact series \eqref{n(z)} is truncated after $\ell={k_0}$.
The (reduced) isothermal compressibility corresponding to the approximation \eqref{A6a} is
\bal
\chi_T(T,z) =& k_B T \frac{\partial n}{\partial p }  =n^{-1}z\partial_z n\nn
 =&n^{-1}\sum_{\ell=1}^{k_0} \ell{c_\ell(T)}\nut^\ell=1+n^{-1}\sum_{\ell=2}^{k_0}(\ell-1){c_\ell(T)}\nut^\ell.
 \label{A6b}
 \eal
Noting again that  $\partial_\mu p=\beta z \partial_z p=n$, the pressure reads
\beq
\beta p(T,z)=\sum_{\ell=1}^{k_0} \frac{c_\ell(T)}{\ell}\nut^\ell.
\label{A8}
\eeq
Taking into account that $\beta\mu^\id=\ln \left[n\Lambda^d(\beta)\right]$, where $\Lambda(\beta)$ is the thermal de Broglie wavelength and $d$ is the dimensionality of the system, it follows that $c_1=\Lambda^{-d}$. The remaining coefficients $\{c_2,c_3,\ldots,c_{k_0}\}$ are directly related to the virial coefficients $\{B_2,B_3,\ldots, B_{k_0}\}$. For instance, $c_2/c_1^2=-2B_2$, $c_3/c_1^3=\frac{3}{2}\left(4B_2^2-B_3\right)$, $c_4/c_1^4=-\frac{4}{3}\left(16B_2^3-9B_2B_3+B_4\right)$, \ldots.

It remains to find the positive real root of the polynomial \eqref{A6a}  as a function of the density $n$ and insert the result into Eq.\ \eqref{A8} to obtain the equation of state. This can be done analytically only if ${k_0}\leq 4$. In particular, in the case ${k_0}=2$ (neglect of triplet correlations) one simply has
\begin{subequations}
\beq
\chi_T(n,T)=2-\frac{1-\sqrt{1-8B_2(T)n}}{4B_2(T) n},
\label{chih3}
\eeq
\beq
Z(n,T)\equiv \frac{p}{nk_BT}=\frac{1}{2}+\frac{1-\sqrt{1-8B_2(T)n}}{8B_2(T) n}.
\label{EOS}
\eeq
\end{subequations}
The equation of state \eqref{EOS} is real-valued only if $n\leq 1/8B_2(T)$. At $n=1/8B_2(T)$ one has $Z=\frac{3}{2}$ and $\chi_T=0$. {}From Eq.\ \eqref{EOS} one can obtain  higher virial coefficients  in terms of $B_2$ as
\beq
B_j=\frac{(2j-3)!!}{j!}2^{2j-3}B_2^{j-1},\quad j\geq 2.
\eeq
In the more general case, ${k_0}\geq 3$, Equations \eqref{A6a} and \eqref{A8} allow us to estimate $B_{{k_0}+1}$ and higher virial coefficients from the knowledge of $\{B_2,B_3,\ldots, B_{k_0}\}$.

\begin{table*}
  \tbl{Comparison between the first few values of the ratio $B_{{k_0}+1}/B_2^{k_0}$ for hard spheres as obtained from the approximation $h_{{k_0}+1}=0$ and the exact values \protect\cite{K82,CM04a,LKM05,CM05,CM06,W13,ZP14,SK14}.}
{\begin{tabular}{@{}ccc}\toprule
   ${k_0}+1$  & $h_{{k_0}+1}=0$&Exact \\
\colrule
  $3$&$4$&$0.62500$\\
 $4$&$-10.375$&$0.28695$\\
 $5$&$33.985$&$0.11025$\\
 $6$&$-109.15$&$0.03888$\\
 $7$&$356.96$&$0.01302$\\
 $8$&$-1181.3$&$0.00418$\\
   \botrule
  \end{tabular}}
  \label{table1}
\end{table*}

\begin{figure}
\begin{center}
\includegraphics[width=8cm]{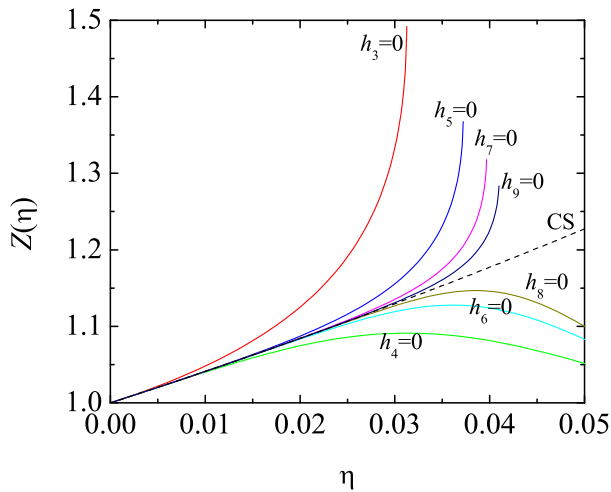}
\caption{Compressibility factor $Z$ as a function of the packing fraction $\eta$ for (three-dimensional) hard spheres as obtained from Eq.\ \protect\eqref{16} with the closures $h_3=0$, \ldots, $h_8=0$. The dashed line represents the Carnahan--Starling equation of state \cite{HM06,CS69}.}
\label{fig1}
\end{center}
\end{figure}

As a test of the reliability of the equation of state, let us study the prediction of $B_{{k_0}+1}$ for (three-dimensional) hard spheres as obtained from the approximation $h_{{k_0}+1}=0$ when the \emph{exact} values \cite{K82,CM04a,LKM05,CM05,CM06,W13,ZP14,SK14} of $\{B_2,B_3,\ldots,B_{k_0}\}$ are used. The results for ${k_0}+1=3$--$8$ are displayed and compared with the exact coefficients $B_{{k_0}+1}$ in Table \ref{table1}.
As  can be seen, neglecting $h_{{k_0}+1}$ in Eq.\ \eqref{16} and imposing the  virial coefficients $\{B_2, B_3,\ldots,B_{k_0}\}$ yield an extremely poor prediction of the next virial coefficient $B_{{k_0}+1}$. The predicted coefficients alternate in sign and are progressively much larger (in absolute value) than the known exact ones.
The poor performance of the approximation defined by Equations \eqref{A6a} and \eqref{A8} for hard spheres is confirmed by Figure \ref{fig1}, where the equation of state is plotted for a sequence of values of ${k_0}+1$. The results suggest that the sequence of approximations $\{h_3=0,h_4=0,\ldots\}$ has an asymptotic character only. It is interesting to note that the curves corresponding to the approximations $h_{{k_0}+1}=0$ with ${k_0}+1=\text{odd}$ terminate with a branch point (where $\chi_T=0$) at certain values of the packing fraction ($\eta=\frac{1}{32}\simeq 0.0313$, $0.0372$, $0.0397$ and $0.0410$ for $h_3=0$, $h_5=0$, $h_7=0$ and $h_9=0$, respectively).

We now depart from the hard-sphere case and assume that the potential energy function $\Phi_N$ is such that a liquid--vapour critical point exists at a certain temperature $T=T_c$ and density $n=n_c$. A simple argument shows that an approximate theory which neglects completely correlations beyond some fixed order $k_0$ loses the possibility of describing such a critical point. The proof proceeds by \emph{reductio ad absurdum} as follows.  If a critical point exists, then the approximation \eqref{A6a} yields $n_c=\sum_{\ell=1}^{k_0}c_\ell(T_c)z_c^\ell$, where $z_c$ is the fugacity at the critical point.
Since $n_c=\text{finite}$, one must have $z_c=\text{finite}$ and  $c_\ell(T_c)=\text{finite}$ for $1\leq\ell\leq k_0$. Consequently, according to Eq.\ \eqref{A6b},
$\lim_{n\to n_c,T\to T_c}\chi_T=n_c^{-1}\sum_{\ell=1}^{k_0}\ell c_\ell(T_c)z_c^\ell=\text{finite}$. But this is in contradiction with the existence of a critical point where the isothermal compressibility diverges.

\section{Multi-particle correlations near the critical point. Generalised critical exponents $\delta_k$}
\label{sec4}
Once we have shown the relevance of keeping multi-particle correlations in the generalised compressibility Equation \eqref{16}, let us turn to the main topic of this work, namely the  behaviour of the integrals $I_k$  near the critical point.

We suppose again that  a liquid--vapour critical point exists at $T=T_c$ and $n=n_c$. The singular behaviour of the relevant thermodynamic functions at criticality is characterised by the so-called \emph{critical exponents} \cite{S71}. In particular, the critical exponent $\delta$ measures how the pressure $p$ tends to its critical value $p_c$ as the density $n$ approaches its critical value $n_c$ along the critical isotherm $T=T_c$:
\beq
\frac{p-p_c}{n_ck_BT_c}\approx \left(\pm 1\right) B\left|\frac{n}{n_c}-1\right|^{\delta},\quad T=T_c,
\label{5a}
\eeq
where $B$ is a dimensionless critical amplitude and $\left(\pm 1\right)\equiv\text{sgn}(n-n_c)$, i.e., $+$ and $-$ correspond to the branches $n>n_c$ and $n<n_c$ of the critical isotherm, respectively. Since the critical isotherm must present an inflection point at the critical point, that is, $\left.\partial^2 p/\partial n^2\right|_c=0$, one must have $\delta>2$.
The behaviour \eqref{5a} implies that the (reduced) isothermal compressibility $\chi_T$  diverges at the critical point according to the law
\beq
\chi_T\approx {A}\left|\frac{n}{n_c}-1\right|^{-(\delta-1)},\quad T=T_c,
\label{5}
\eeq
where we have called $A\equiv 1/B\delta$.
{}From Equation \eqref{compress_eqn} it follows that
\beq
I_2\equiv n^2\int\dd 2\, h_2(1,2)\approx {A}n_c\left|\frac{n}{n_c}-1\right|^{-(\delta-1)},\quad T=T_c.
\label{5c}
\eeq
The divergence of the integral $I_2$ at the critical point is directly linked to the development of long-range \emph{pair} correlations.

It is physically expected that the clusters of three, four, five, \ldots particles are also highly correlated near the critical point in such a way that the generalised integrals $I_k$ defined by Eq.\ \eqref{1} diverge at criticality. In order to characterise this behaviour on the critical isotherm, let us define the \emph{generalised} critical exponents $\delta_k$ by
\beq
I_k\approx \left(\pm 1\right)^k{A_k}n_c\left|\frac{n}{n_c}-1\right|^{-(\delta_k-1)},\quad T=T_c.
\label{6}
\eeq
In particular, $\delta_2=\delta$ and $A_2=A$. Equation \eqref{6} implies that
\beq
\partial_nI_k\approx -\left(\pm 1\right)^{k+1}(\delta_k-1){A_k}\left|\frac{n}{n_c}-1\right|^{-\delta_k},\quad T=T_c.
\label{6b}
\eeq
Combining this with Eq.\ \eqref{5},
\beq
\chi_Tn\partial_nI_k\approx -\left(\pm 1\right)^{k+1}(\delta_k-1){A_k} An_c\left|\frac{n}{n_c}-1\right|^{-(\delta_k+\delta-1)},\quad T=T_c.
\label{6c}
\eeq
Clearly, $\chi_Tn\partial_nI_k\gg k I_k$ near the critical point. Therefore, insertion of Equations \eqref{6} (with $k\to k+1$) and \eqref{6c} into Eq.\ \eqref{2} yields
\begin{subequations}
\bal
\delta_{k+1}=&\delta_k+\delta,
\label{8}
\\
A_{k+1}=&-AA_k(\delta_k-1).
\label{7}
\eal
\end{subequations}
The solutions to the above recursion relations are
\begin{subequations}
\label{9+9b}
\bal
\delta_k=&(k-1)\delta,
\label{9}
\\
A_k=&(-1)^k A^{k-1}\left(\delta-1\right)\left(2\delta-1\right)\cdots \left[(k-2)\delta-1\right].
\label{9b}
\eal
\end{subequations}

It is instructive to rederive Equations \eqref{9+9b}  from Eq.\ \eqref{16b}. Note first that the thermodynamic relation $n=\beta z\partial_z p$, together with Eq.\ \eqref{5a}, yields
\beq
\frac{n}{n_c}-1\approx \left(\pm 1\right) B^{-1/\delta}\left|\frac{z}{z_c}-1\right|^{1/\delta},\quad T=T_c.
\label{n_c-z_c}
\eeq
This is equivalent to $(p-p_c)/n_ck_BT_c\approx (z/z_c-1)$.
At $T=T_c$ and for $n\approx n_c$ the most divergent term on the right-hand side of Eq.\ \eqref{16b} is $z_c^k\partial_z^k n$ and thus
\bal
I_{k+1}\approx &z_c^k\partial_z^k n\approx
n_c \frac{B^{-1/\delta}}{\delta} \frac{\Gamma\left(k-\delta^{-1}\right)}{\Gamma\left(1-\delta^{-1}\right)}\left|\frac{z}{z_c}-1\right|^{1/\delta-k}\nn
\approx&\left(\pm 1\right)^{k+1}A^k n_c \left(-\delta\right)^{k-1}\frac{\Gamma\left(k-\delta^{-1}\right)}{\Gamma\left(1-\delta^{-1}\right)}\left|\frac{n}{n_c}-1\right|^{-\left(k\delta-1\right)}.
\label{17}
\eal
Comparison between Equations \eqref{6} (with $k\to k+1$) and \eqref{17} yields
$\delta_{k+1}=k\delta$ and $A_{k+1}=A^{k} \left(-\delta\right)^{k-1}{\Gamma\left(k-\delta^{-1}\right)}/{\Gamma\left(1-\delta^{-1}\right)}$,
in agreement with Equations \eqref{9+9b}.

Equations \eqref{9+9b} are the basic results of this paper. Equation \eqref{9} implies that, near the critical point, the correlation integral over all the coordinates of a cluster of $k$ particles diverges with a critical exponent that increases \emph{linearly} with $k$. As shown by Eq.\ \eqref{17}, this linear growth is a direct consequence of the exact formula \eqref{16b}. In addition, according to Eq.\ \eqref{9b}, the critical amplitude alternates in sign and its absolute value grows almost exponentially with $k$. Moreover, the presence of the  factor  $\left(\pm 1\right)^k$ in Eq.\ \eqref{6} means that $I_k$ with $k=\text{odd}$ diverges to either $-\infty$ or $+\infty$ depending on whether the critical point is reached along the critical isotherm with $n>n_c$ or $n<n_c$, respectively.
To understand this peculiar property, let us focus on $I_3$ and note that, according to Eq.\ \eqref{2}, this quantity is related to the curvature of the isotherm by
\beq
I_3=n(\chi_T-1)(\chi_T-2)-\frac{n^2}{k_BT}\chi_T^3\frac{\partial^2 p}{\partial n^2}.
\label{I3}
\eeq
Near the critical point, the first term on the right-hand side of Eq.\ \eqref{I3} is dominated by the second one, i.e.,
\beq
I_3\approx-\frac{n_c^2}{k_BT_c}\chi_T^3\frac{\partial^2 p}{\partial n^2}.
\label{I3b}
\eeq
The second derivative $\partial^2p/\partial n^2$ along the critical isotherm is positive for $n>n_c$ and negative for $n<n_c$ and therefore $I_3$ has the opposite behaviour. In addition, $\partial^2p/\partial n^2\to 0$ and $\chi_T\to\infty$ as given by Equations \eqref{5a} and \eqref{5}, respectively. As a consequence,
\beq
I_3\approx -\left(\pm 1\right) A^2(\delta-1)n_c\left|\frac{n}{n_c}-1\right|^{-(2\delta-1)},\quad T=T_c,
\label{I3c}
\eeq
in consistency with Equations \eqref{6} and \eqref{9+9b}.
\section{Kirkwood's superposition approximation}
\label{sec5}
The standard approximation to eliminate the triplet distribution function in favour of the pair distributions is KSA \cite{H56,C68}. It reads
\beq
g_3(1,2,3)=g_2(1,2)g_2(1,3)g_2(2,3).
\label{KSA}
\eeq
Making use of the cluster relations \eqref{29} and \eqref{30}, it turns out that the KSA is equivalent to
\bal
h_3(1,2,3)=&h_2(1,2)h_2(1,3)+h_2(1,2)h_2(2,3)+h_2(1,3)h_2(2,3)\nn
&+h_2(1,2)h_2(1,3)h_2(2,3).
\label{12}
\eal
Thus, in this approximation
\beq
I_3=3\frac{I_2^2}{n}+n^3\int d2\int d3\, h_2(1,2)h_2(1,3)h_2(2,3).
\label{13}
\eeq

Let us now see that Eq.\ \eqref{13} is incompatible with the existence of a critical point, i.e., the approximation \eqref{13} is incompatible with the exact relation \eqref{2} (with $k=2$) near the critical point \cite{PSK13}.
To that end, we start by assuming that a critical point does exist. If that was the case,  Appendix  \ref{appC} shows that the second term on the right-hand side of Eq.\ \eqref{13}  would be negligible versus the first term near the critical point, yielding
\beq
I_3\approx 3\frac{I_2^2}{n_c}
\approx 3A^2n_c\left|\frac{n}{n_c}-1\right|^{-2(\delta-1)},\quad T=T_c.
\label{13b}
\eeq
This is in contradiction with the exact relation \eqref{I3c} at least in two important points: (1) the divergence of $I_3$  is weaker (by a difference of unity in the exponent) in Eq.\ \eqref{13b} than in Eq.\ \eqref{I3c} and (2) the sign of $I_3$ in Eq.\ \eqref{13b} is always positive regardless of whether $n<n_c$ or $n>n_c$, in contrast to Equations \eqref{I3b} and \eqref{I3c}. In other words, the true long-range nature of the triplet correlation function $h_3$ is stronger than the one induced by that of $h_2$ in the KSA \eqref{12}.

We can reformulate the inconsistency between Equations \eqref{I3c} and \eqref{13b} as follows. Let us define a residual, non-KSA term as
\bal
\Delta h_3(1,2,3)=&h_3(1,2,3)-\left[h_2(1,2)h_2(1,3)+h_2(1,2)h_2(2,3)+h_2(1,3)h_2(2,3)\right.\nn
&\left.+h_2(1,2)h_2(1,3)h_2(2,3)\right].
\label{Delta_h3}
\eal
Then, near the critical point,
\beq
J_3\equiv n^3\int \dd 2\int \dd 3\,\Delta h_3(1,2,3)\approx I_3-3\frac{I_2^2}{n_c}.
\eeq
Since $I_2^2/I_3\sim |n/n_c-1|\to 0$, it follows that $|J_3|\gg I_2^2/n_c$.
This means that the integral $J_3$ of the residual function $\Delta h_3$ diverges near the critical point more rapidly than $I_2^2$ and
therefore cannot be neglected (as the KSA does) near the critical point.

Thus, we conclude that the KSA is inconsistent with the existence of a critical point. The \emph{no-go} arguments of this section are essentially equivalent to the proof presented in Ref.\ \cite{PSK13}.

\section{Generalised superposition approximations}
\label{sec6}
The KSA \eqref{KSA} has been extended to express the $k$-particle distribution function in terms of pair distributions \cite{CG84,RS85a,RS85b}:
\beq
g_k(1,2,\ldots,k)=\prod_{1\leq i<j\leq k} g_2(i,j).
\label{25}
\eeq
However, this cannot be considered as a true GSA closure since in the latter $g_k$ must be expressed in terms of $g_2$, \ldots, $g_{k-1}$ and not only of $g_2$ \cite{C59,C68}. For instance, a GSA at the level of $g_4$ is \cite{C68,S04}
\beq
g_4(1,2,3,4)=\frac{g_3(1,2,3)g_3(1,2,4)g_3(1,3,4)g_3(2,3,4)}{g_2(1,2)g_2(1,3)g_2(1,4)g_2(2,3)g_2(2,4)g_2(3,4)}.
\label{GSA}
\eeq
Inserting Eqs. \eqref{29}--\eqref{31} into Eq.\ \eqref{GSA} it is possible to express $h_4$ in terms of $h_3$ and $h_2$. The full expression is rather long and will be omitted here. It has the structure
\bal
h_4(1,2,3,4)=&\frac{1}{D_4(1,2,3,4)}\left\{\textbf{12}h_2(1,2)h_3(1,3,4)+\textbf{6}h_3(1,2,3)h_3(1,3,4)\right.\nn
&\left.-\textbf{12}h_2(1,2)h_2(1,3)h_2(2,4)-2\left[\textbf{4}h_2(1,2)h_2(1,3)h_2(1,4)\right]+\cdots\right\},
\label{h4GSA}
\eal
where
\beq
D_4(1,2,3,4)\equiv g_2(1,2)g_2(1,3)g_2(1,4)g_2(2,3)g_2(2,4)g_2(3,4).
\label{D4}
\eeq
In Eq.\ \eqref{h4GSA} the ellipsis represents terms expressed as products by (at least) a factor $h_2$ or $h_3$ of terms similar to the indicated ones. Moreover, as in Equations \eqref{30} and \eqref{31}, a numerical coefficient in boldface represents the number of equivalent terms obtaining by  relabelling of $1$--$4$.  The origin of these combinatorial coefficients  just lies in the fact that the
correlation function $h_4(1,2,3,4)$ is a symmetric function of its arguments.

To proceed, we need to resort to heuristic arguments, similar to the ones followed in Appendix  \ref{appC}. Assuming that a critical point does exist, the integral $I_4$ near the critical point is dominated by configurations where particles $1$--$4$ are widely separated from each other. In those configurations, $h_2\to 0$ (even if it decays algebraically) and therefore we can replace $g_2\to 1$ in Equation \eqref{D4} and set $D_4\to 1$ when integrating both sides of Equation \eqref{h4GSA}. By the same argument, the terms represented by the ellipsis in Equation \eqref{h4GSA} can be neglected versus the displayed ones. Therefore,
\bal
I_4\approx &n^4\int \dd2\int\dd3\int\dd4\left[12h_2(1,2)h_3(1,3,4)+6h_3(1,2,3)h_3(1,3,4)\right.\nn
&\left.-12h_2(1,2)h_2(1,3)h_2(2,4)-8h_2(1,2)h_2(1,3)h_2(1,4)\right],
\label{I4GSA}
\eal
where we have taken into account that all the equivalent terms give the same integral.
It follows from Baxter's Equation (7) that the term $\int\dd2\int\dd4\, h_2(1,2)h_3(1,3,4)$ is linear in $h_2(1,3)$ whereas the integral $\int\dd2\int\dd4\, h_3(1,2,3)h_3(1,3,4)$ yields an expression quadratic in $h_2(1,3)$,  thus decaying faster than {the} former one when $r_{13}\to\infty$.
Therefore, the second term in the integrand of Equation \eqref{I4GSA} can be neglected versus the first term and thus
\beq
I_4\approx 4I_2\frac{3I_3-5I_2^2/n_c}{n_c}.
\eeq
As a final step, we note that, as discussed in Section \ref{sec5}, $|I_3|\gg I_2^2/n_c$ near the critical point, so that
\beq
I_4\approx 12\frac{I_2I_3}{n_c}\approx-\left(\pm 1\right)12 A^3(\delta-1)n_c\left|\frac{n}{n_c}-1\right|^{-(3\delta-2)},\quad T=T_c,
\label{I4c}
\eeq
where use has been made of Equations \eqref{5c} and \eqref{I3c}. Equation \eqref{I4c} is the GSA counterpart of the KSA relation \eqref{13b}. Analogously to the latter case, Equation \eqref{I4c} is inconsistent with Equations \eqref{9+9b}, according to which $\delta_4-1=3\delta-1>3\delta-2$ and $A_4=A^3(\delta-1)(2\delta-1)\neq -12\left(\pm 1\right) A^3(\delta-1)$. Thus, we conclude that the GSA closure \eqref{GSA} is also incompatible with the existence of a critical point.

Suppose now a generalisation of both \eqref{13b} and \eqref{I4c} where the correlation integral $I_{k_0+1}$ (with $k_0\geq 4$) is approximated near the (potential) critical point by
\beq
I_{k_0+1}\sim \prod_{j=2}^{k_0} I_j^{x_j},
\label{36}
\eeq
where $x_j$ are non-negative integer exponents satisfying the constraint
\beq
\sum_{j=2}^{k_0} j x_j={k_0}+2.
\label{sum_x}
\eeq
For instance, ${k_0}=2$ and $x_2=2$ in Equation \eqref{13b}, while ${k_0}=3$ and $x_2=x_3=1$ in Equation \eqref{I4c}. The constraint \eqref{sum_x} implies that
\beq
\sum_{j=2}^{k_0} x_j\geq 2,
 \label{sum_x'}
\eeq
since $\sum_{j=2}^{k_0} x_j=1$ is possible only if $x_j=\delta_{jj_0}^{\text{Kr}}$ with some $2\leq j_0\leq {k_0}$ (where  $\delta_{ij}^{\text{Kr}}$ is the Kronecker symbol), thus violating Equation \eqref{sum_x}.

According to Equations \eqref{6} and \eqref{36},
\beq
\delta_{{k_0}+1}-1=\sum_{j=2}^{{k_0}} x_j(\delta_j-1).
\label{37}
\eeq
Using Equation \eqref{9} for $\delta_j$ with $2\leq j\leq {k_0}$, one has
\bal
\delta_{{k_0}+1}&=1+({k_0}+2)\delta-(\delta+1)\sum_{j=2}^{{k_0}}x_j\nn
&\leq {k_0}\delta-1.
\label{38}
\eal
In the first equality use has been made of Equation \eqref{sum_x}, while in the second step the inequality \eqref{sum_x'} has been used.
As in the cases of Equations \eqref{13b} and \eqref{I4c}, the GSA \eqref{36} violates the exact condition $\delta_{{k_0}+1}={k_0}\delta$ by at least a unity.

Therefore, one can conclude that neglecting residual correlations  beyond a certain finite order in a GSA eliminates the critical point from
the resulting approximate theory. Stated in an alternative way,
a necessary condition to describe a critical isotherm is to take into account correlations of
arbitrary order not accounted for by lower order correlations.

\section{Concluding comments}
\label{sec7}
Although the hierarchy \eqref{0} contains most fundamental information about equilibrium correlations, it has been rarely
used up to now. Our study shows clearly its predicting power as far as the description of critical phenomena is concerned.
In an integrated form, the hierarchy permits to express the spatial integral of the $(k+1)$-particle correlation function in terms
of thermodynamic derivatives involving the isothermal compressibility, as shown by Equation \eqref{16}. The study of these generalised compressibility
equations lead to the following conclusions
\begin{enumerate}
\item
Correlations of any order contribute in an essential way to the occurrence of a critical point. In other words,
approximations which neglect correlations beyond some fixed order cannot predict critical behaviour.
\item
If a fluid undergoes a liquid--vapour phase transition, the critical exponents characterising the divergence
of spatial integrals of $(k+1)$-particle correlation functions grow linearly with $k$, Equation  \eqref{9}.
\item
Not only the KSA but also its generalisations are inconsistent with
the existence of a critical point.
\end{enumerate}

An open problem suggested by these results is the validity of the conjecture stating that any theory which
expresses three-particle correlations in terms of two-particle ones loses the possibility of describing a critical point (a specific example is here the KSA). A stronger conjecture would claim that once the higher order correlations are assumed to be functionals of the lower order ones the resulting theory becomes inconsistent with a critical behaviour.

As {an additional} comment, let us remark that, from a strict point of view, the inconsistency of the KSA with the existence of a critical point follows when it is used as a closure for the Baxter hierarchy \eqref{0b}. This leaves a question on whether the KSA would lead to the same conclusion when applied as a closure to the Yvon--Born--Green  hierarchy \cite{HM06,H56}. However, numerical results combined with analytical techniques \cite{JKLFF81,FF81b,F81,FF83,JLK83,PSK13b} suggest that also in this case a true critical behaviour does not appear for three-dimensional systems. A challenging theoretical problem is to work out an exact analytic answer also to this question.

{Finally, it must be noted that we focused
entirely on the effect of multi-particle correlations on the existence of a critical point via the critical exponent $\delta$. This is because the generalised compressibility Equations \eqref{16} are directly related to the isothermal compressibility and its derivatives with respect to density at constant temperature. Therefore, the problems associated with
other critical exponents different from $\delta$ have not been studied.
  Of course, if the conditions for criticality are not satisfied, we
should not expect any singularity not only in the isothermal
compressibility but also in other thermodynamic quantities, such as, for instance,
in the specific heat along the critical isochore (as measured by the critical exponent $\alpha$).}

\section*{Acknowledgements}
Andr\'es Santos is grateful to J.J. Ruiz-Lorenzo for insightful discussions.

\section*{Disclosure statement}
No potential conflict of interest was reported by the authors.

\section*{Funding}
The research of Andr\'es Santos has been partially financed by the Spanish Government [grant number FIS2013-42840-P]; the Regional Government of Extremadura (Spain) [grant number GR15104] (partially financed by the ERDF).


\appendices

\section{Right-hand side of Equation \protect\eqref{13} near the critical point}
\label{appC}
In this Appendix we show by heuristic arguments that the second term on the right-hand side of Equation \protect\eqref{13} can be neglected versus the first term near the critical point. To that end, let us compare the integrals
\begin{subequations}
\label{C3}
\bal
K\equiv &\int \dd2\int \dd3\, h_2(1,2)h_2(1,3)h_2(2,3),
\label{C1}
\\
\frac{I_2^2}{n^4}=&\int \dd2\int \dd3\, h_2(1,2)h_2(1,3).
\label{C2}
\eal
\end{subequations}

The factor $h_2(2,3)$ in the integrand of Equation \eqref{C1} is replaced by unity in Equation \eqref{C2}.
Near the critical point, where the pair correlations become long-ranged, both integrals are dominated by configurations where particles $1$, $2$ and $3$ are widely separated from each other. In those configurations, even if $h_2(2,3)$ decays algebraically with $r_{23}$, the replacement $h_2(2,3)\to 1$ when going from $K$ to $I_2^2/n^4$ makes $K$ diverge much more slowly than $I_2^2/n^4$.

To refine the above qualitative argument, note that Equations \eqref{C3} can be rewritten as
\begin{subequations}
\label{C4}
\bal
K=&\frac{1}{2\pi}\int \dd \bm{q}\,\left[\widetilde{h}_2(\bm{q})\right]^3,
\label{C5}
\\
\frac{I_2^2}{n^4}=&\left[\widetilde{h}_2(0)\right]^2
\label{C6}
\eal
\end{subequations}
where
\beq
\widetilde{h}_2(\bm{q})=\int \dd \bm{r}\, h_2(\bm{r})\ee^{\ii\bm{q}\cdot\bm{r}}
\eeq
is the Fourier transform of the pair correlation function.
As a toy function, let us consider a three-dimensional geometry and the (classical) asymptotic form
\beq
h_2(\bm{r})\sim \frac{\ee^{-\kappa r}}{r}, \quad \kappa r\gg 1,
\label{C7}
\eeq
where $\kappa$ is the inverse correlation length, which goes to zero as the critical point is approached. The Fourier transform of \eqref{C7} is
\beq
\widetilde{h}_2(\bm{q})\sim \left(\kappa^2+q^2\right)^{-1}, \quad q/\kappa \ll 1.
\label{C8}
\eeq
Thus, $I_2^2$ near the critical point diverges as
\beq
I_2^2\sim \kappa^{-4}.
\label{C9}
\eeq
In contrast, the integral $K$ behaves as
\bal
K\sim &\int \dd \bm{q}\, \left(\kappa^2+q^2\right)^{-3}=\kappa^{-3} \int \dd \bm{q}^*\, \left(1+{q^*}^2\right)^{-3}\nn
\sim &\kappa^{-3}.
\eal
Therefore, $K/(I_2^2/n^4)\to 0$, as expected.

\end{document}